\documentclass[prl,showpacs,floatfix,twocolumn]{revtex4-1}
\usepackage{amsfonts}
\usepackage{stmaryrd}
\usepackage{bbm}
\usepackage{mathrsfs}
\usepackage{tipa}
\usepackage{amssymb}
\usepackage{txfonts}
\usepackage{graphicx}
\usepackage{dcolumn}
\usepackage{epstopdf}
\usepackage[colorlinks,linkcolor=blue,urlcolor=blue,citecolor=blue]{hyperref}
\newcommand{\PreserveBackslash}[1]{\let\temp=\\#1\let\\=\temp}
\newcolumntype{C}[1]{>{\PreserveBackslash\centering}p{#1}}
\newcolumntype{R}[1]{>{\PreserveBackslash\raggedleft}p{#1}}
\newcolumntype{L}[1]{>{\PreserveBackslash\raggedright}p{#1}}

\begin{document}

\newcommand*{\cm}{cm$^{-1}$}

\title{Correlation-induced self-doping in intercalated iron-pnictide superconductor Ba$_2$Ti$_2$Fe$_2$As$_4$O}

\author{J.-Z.  \surname{Ma$^1$}}
\thanks{These authors contribute equally to this work.}
\author{A. van \surname{Roekeghem$^{1,2}$}} 
\thanks{These authors contribute equally to this work.}
\author{P. \surname{Richard$^{1,3}$}}
\author{Z.-H. \surname{Liu$^{4}$}}
\author{H. \surname{Miao$^{1}$}}
\author{L.-K. \surname{Zeng$^{1}$}}
\author{N. \surname{Xu$^{5}$}}
\author{M. \surname{Shi$^{5}$}}
\author{C. \surname{Cao$^{6}$}}
\author{J.-B. \surname{He$^{4}$}}
\author{G.-F. \surname{Chen$^{1,3,4}$}}
\author{Y.-L. \surname{Sun$^{7}$}}
\author{G.-H. \surname{Cao$^{7}$}}
\author{S.-C. \surname{Wang$^{4}$}}
\author{S. \surname{Biermann$^{2,8,9}$}}
\author{T.  \surname{Qian$^{1}$}}\email{tqian@iphy.ac.cn}
\author{H. \surname{Ding$^{1,3}$}}\email{dingh@iphy.ac.cn}
\affiliation{$^1$ Beijing National Laboratory for Condensed Matter Physics, and Institute of Physics, Chinese Academy of Sciences, Beijing 100190, China}
\affiliation{$^2$ Centre de Physique Th\'{e}Žorique, Ecole Polytechnique, CNRS-UMR7644, 91128 Palaiseau, France}
\affiliation{$^3$ Collaborative Innovation Center of Quantum Matter, Beijing, China}
\affiliation{$^4$ Department of Physics, Renmin University, Beijing, 100872, China}
\affiliation{$^5$ Paul Scherrer Institute, Swiss Light Source, CH-5232 Villigen PSI, Switzerland}
\affiliation{$^6$ Condensed Matter Physics Group, Department of Physics,Hangzhou Normal University, Hangzhou 310036, China}
\affiliation{$^7$ Department of Physics, Zhejiang University, Hangzhou 310027, China}
\affiliation{$^8$ Coll$\grave{e}$ge de France, 11 place Marcelin Berthelot, 75005 Paris, France}
\affiliation{$^9$ European Theoretical Synchrotron Facility (ETSF), Europe}


\begin{abstract}
The electronic structure of the intercalated iron-based superconductor Ba$_2$Ti$_2$Fe$_2$As$_4$O (\emph{T}$_{c}$ $\sim$ 21.5 K) has been investigated by using angle-resolved photoemission spectroscopy and combined local density approximation and dynamical mean field theory calculations. The electronic states near the Fermi level are dominated by both the Fe 3$d$ and Ti 3$d$ orbitals, indicating that the spacing layers separating different FeAs layers are also metallic. By counting the enclosed volumes of the Fermi surface sheets, we observe a large self-doping effect, $i.e.$ 0.25 electrons per unit cell are transferred from the FeAs layer to the Ti$_2$As$_2$O layer, leaving the FeAs layer in a hole-doped state. This exotic behavior is successfully reproduced by our dynamical mean field calculations, in which the self-doping effect is attributed to the electronic correlations in the Fe 3\emph{d} shell. Our work provides an alternative route of effective doping without element substitution for iron-based superconductors.
\end{abstract}

\pacs{71.27.+a, 73.20.-r, 74.70.Xa, 79.60.-i}

\keywords{angle-resolved photoemission spectroscopy,self doping, FeAs, iron chalcogenides, superconductor}
\maketitle


In iron-based superconductors (IBSCs), the most common ways to suppress long-range antiferromagnetic order and obtain high-$T_c$ superconductivity is to introduce carriers \cite{Hosono,BKFA,BFCA} and/or internal strain \cite{BFAP,BFRA,SFRA} by element substitution. However, an inevitable problem is that element substitution also introduces disorder, and impurity scattering is believed to be detrimental to superconductivity \cite{impurity}, though not as seriously as in cuprate superconductors. It has been revealed that the impurity scattering effects are site dependent and the scattering strength is gradually reduced when the dopants move away from the Fe plane \cite{Feng,Uchida}. This may partially explain why the maximum $T_c$ is much higher and the superconducting dome is much wider in (Ba,K)Fe$_2$As$_2$ as compared with Ba(Fe,Co)$_2$As$_2$ \cite{Feng}. Therefore, finding an alternative way to dope carriers but without introducing disorder would be a promising path for reaching higher $T_c$ superconductivity. 

A remarkable feature in the IBSCs is that there is an intimate relationship between the electronic correlations and the \emph{d}-shell occupancy. For hole doping, extremely low coherence temperatures are expected, while electron doping reinforces Fermi-liquid properties \cite{Werner NPhys,Xu PRX}. Moreover, the orbital polarization can be tuned by the magnitudes of the Coulomb interaction \emph{U} and the Hund's rule coupling \emph{J}, leading to a redistribution of electrons among five Fe 3\emph{d} orbitals \cite{ZFang RRL}. Electronic correlations weaken the hybridization between Fe and ligand atoms, reducing the effective occupancy of the Fe 3$d$ orbitals \cite{DMFT FeSe}. However, as the total electron count on the Fe and ligand atoms in crystals, such as BaFe$_{2}$As$_{2}$, is conserved, such charge redistribution between them does not produce any doping effect on the Fermi surfaces (FSs).

\begin{figure}[ht]
\centering
\includegraphics[scale=0.48]{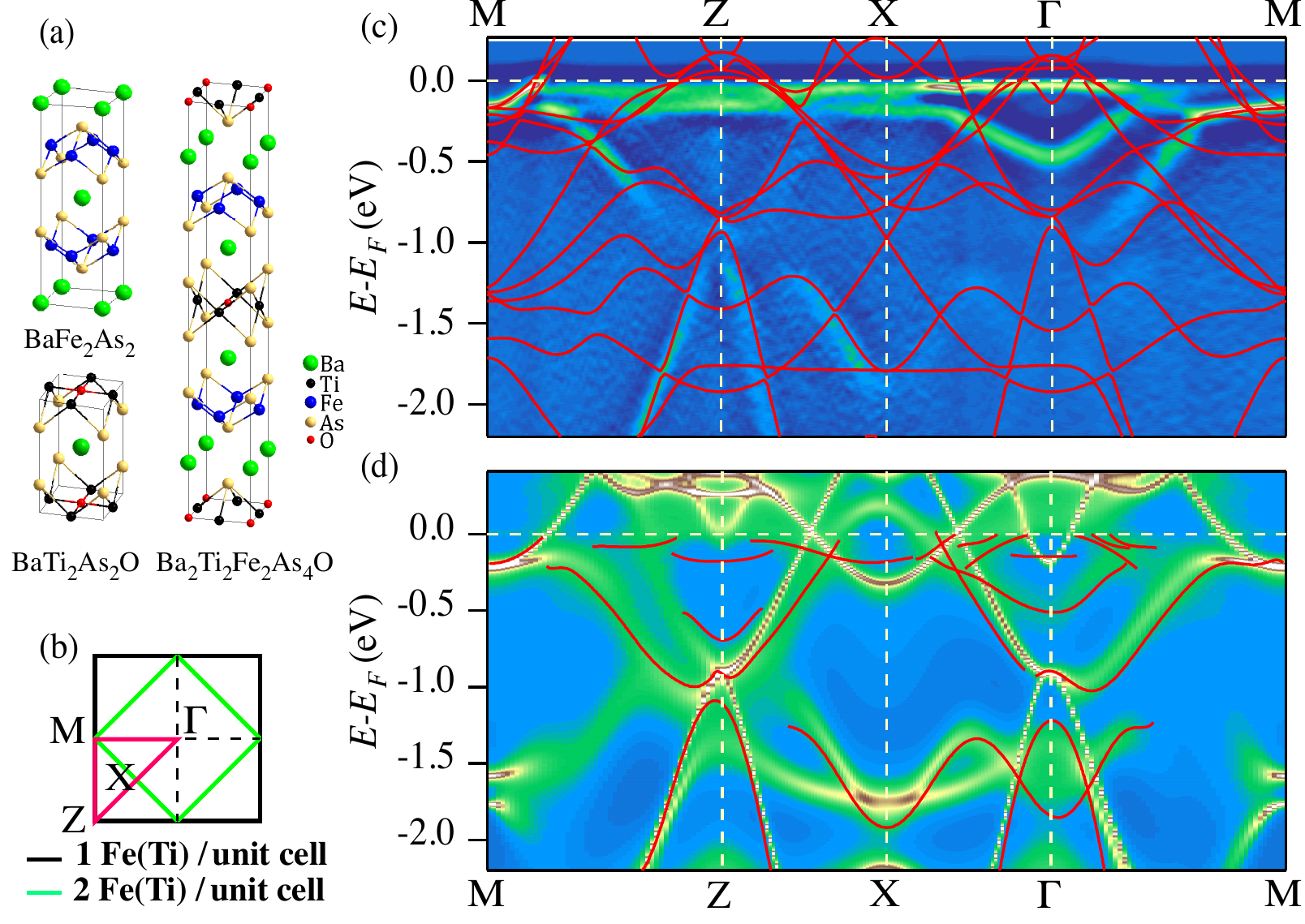}
\caption{\label{fig:one}(Color online) (a) Crystal structures of BaFe$_{2}$As$_{2}$, BaTi$_{2}$As$_{2}$O and Ba$_{2}$Ti$_{2}$Fe$_{2}$As$_{4}$O. (b) Schematic Brillouin zones for one and two Fe (Ti) per unit cell. Red lines indicate the measurement locations in panel (c). (c) Two-dimensional (2D) curvature intensity plot \cite{PZhang_RSI} of the ARPES data along MZX$\Gamma$M recorded at 150 K and photon energy $h\nu$ = 55 eV with HP. The LDA bands are also plotted without renormalization for comparison. (d) Momentum-resolved spectral function calculated within LDA + DMFT at $T$ = 145 K. Red curves represent the extracted experimental band dispersions.}
\end{figure}

In this work, we prove that doping can be induced by electronic correlations in the IBSC Ba$_2$Ti$_2$Fe$_2$As$_4$O (Ba22241, \emph{T}$_{c}$ $\sim$ 21.5 K) due to the intercalation of metallic Ti$_2$As$_2$O layers. Ba22241 can be regarded as a superlattice consisting of alternating stacking of BaFe$_2$As$_2$ and BaTi$_2$As$_2$O layers [Fig. \ref{fig:one}(a)] \cite{Cao JACS}. Compared with other IBSCs, the most distinctive characteristic of Ba22241 is the metallic nature of the intercalated layers, which contributes distinctly to the density of states (DOS) at the Fermi level (\emph{E}$_{F}$) \cite{Cao LDA}. Our ARPES measurements suggest that the low-energy band dispersions can be regarded as a superposition of the band structures of the FeAs and Ti$_2$As$_2$O layers. By counting the volumes of the FSs, we find that about 0.25 electrons per unit cell are transferred from the FeAs layer to the Ti$_2$As$_2$O layer. This exotic behavior is successfully reproduced by dynamical mean field-based electronic structure calculations, which allow us to identify electronic Coulomb correlations in the 3\emph{d} shells as the main cause for this self-doping effect.

High-quality single crystals of Ba22241 were synthesized by the flux method \cite{Cao Crystal}. ARPES measurements were performed at beamlines PGM and APPLE-PGM of the Synchrotron Radiation Center (Wisconsin) with Scienta R4000 and SES 200 analyzers, respectively, as well as at beamline SIS of the Swiss Light Source (PSI) with Scienta R4000. The energy and angular resolutions were set at 15-30 meV and 0.2$^{\circ}$, respectively. The samples were cleaved \emph{in situ} and measured in the temperature range between 25 and 150 K in a vacuum better than 3 $\times$ 10$^{-11}$ Torr. The ARPES data were taken with vertical exit slits under horizontal (HP) or vertical (VP) polarized lights at SRC and with horizontal exit slit at PSI.

Figure \ref{fig:one}(c) shows the experimental band dispersions along the high-symmetry lines MZX$\Gamma$M in an energy range within 2.2 eV below $E_F$. The band dispersions are much more complex than those of other IBSCs due to the contribution of the metallic Ti$_{2}$As$_{2}$O layers. To understand the multiband electronic structure, we superimpose the Kohn-Sham band structure of density functional theory within the local density approximation (LDA) as calculated within Wien2k \cite{Wien2k} on top of the experimental data in Fig. \ref{fig:one}(c). We use the experimental crystal structure of Ref. \cite{Cao JACS} as input. The experimental band dispersions below --1 eV match well the LDA bands, which are mainly derived from As 4$p$ of the Ti$_{2}$As$_{2}$O layers [Fig. S3 in the Supplement Materials]. Obvious discrepancies between the experimental results and the LDA calculations are observed within 1 eV below $E_F$, where the DOS is mainly of Fe $3d$ and Ti $3d$ characters. The deviation can be qualitatively attributed to non-negligible correlation effects between the 3$d$ electrons, for which the self-energy leads to strong band renormalizations near $E_F$, as observed in other IBSCs \cite{Pierre review}. To analyze the effects of Coulomb correlations on the electronic structure, we have performed LDA + dynamical mean field theory (DMFT) calculations of the momentum-resolved spectral function, starting from the above Kohn-Sham band structure and the implementation of Ref. \cite{Aichhorn 2009} using projected atomic orbitals. Since our target compound contains two different atomic species with partially filled narrow $d$-shells, we have generalized the usual LDA + DMFT scheme to include effective local Coulomb interactions on both the Fe-$3d$ and Ti-$3d$ shells. We use $U$ = 2.64 eV [3.50 eV] and $J$ = 0.96 eV [0.74 eV] for the Hubbard interactions (monopole Slater integral F$_0$ and Hund's coupling on Fe [Ti]), and a multi-orbital around mean field double counting based on the LDA electron count \cite{AVR BaCo2As2}. The resulting 10-orbital many-body problem was solved within a continuous-time Quantum Monte Carlo scheme, as implemented in the TRIQS toolbox \cite{TRIQS}. The corresponding results are presented in Fig. \ref{fig:one}(d), which reproduce well the experimental band dispersions. The Fe 3$d$ derived bands are strongly renormalized compared with the LDA results and their spectral intensities are very diffuse due to a large quasiparticle broadening encoded in the imaginary part of the DMFT self energy.

\begin{figure}[ht]
\centering
\includegraphics[scale=0.55]{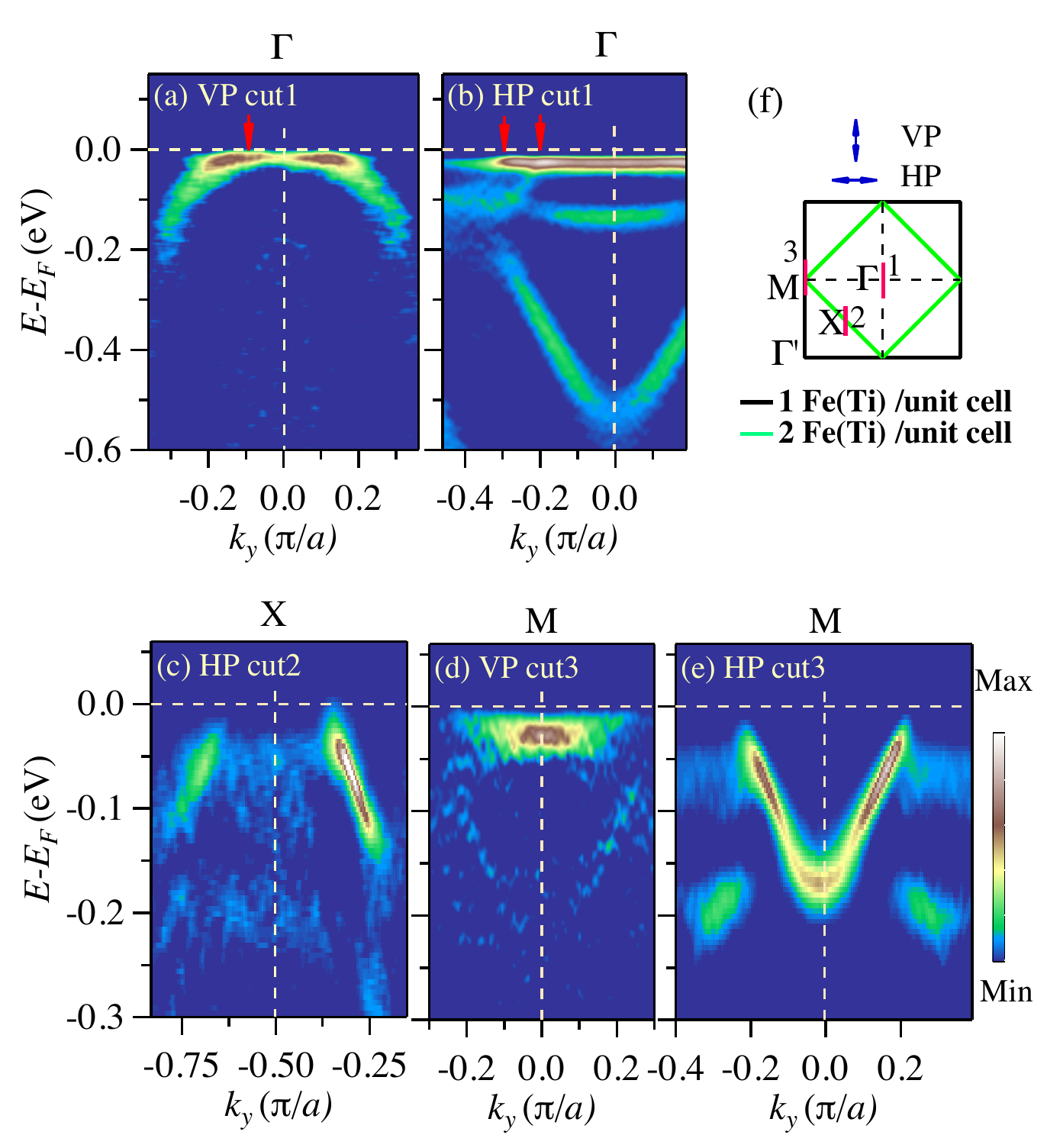}
\caption{\label{fig:two}(Color online) (a-e) 2D curvature intensity plots of the near-$E_F$ ARPES data recorded at $h\nu$ = 55 eV. The momentum locations are indicated as red vertical lines in panel (f). To illustrate the Fe-3$d$ related band dispersions more clearly, the data taken at 30 K are shown in panels (a), (b) and (d). As the Ti-3$d$ related band dispersions are dramatically changed across the transition at 125 K [Fig. S2 in the Supplement Materials], the data taken at 150 K are shown in panels (c) and (e).}
\end{figure}

A prominent feature in the calculations is that the intercalated Ti$_2$As$_2$O layers contribute significantly to the DOS at $E_F$, demonstrating the metallic nature of the intercalated layers in this system. To clarify the effects on the low-energy electronic states of the FeAs layers, we show the band dispersions near $E_F$ in Fig. \ref{fig:two}. All the cuts are parallel to the $\Gamma$M direction, as schematically plotted in Fig. \ref{fig:two}(f). We identify three hole-like bands near $\Gamma$ [Figs. \ref{fig:two}(a) and \ref{fig:two}(b)], one hole-like band near X [Fig. \ref{fig:two}(c)] and two electron-like bands near M [Figs. \ref{fig:two}(d) and \ref{fig:two}(e)], which cross $E_F$. The band dispersions can be regarded as a superposition of the band structures of the FeAs and Ti$_{2}$As$_{2}$O layers. We assign the three hole-like bands near $\Gamma$ [Figs. \ref{fig:two}(a) and \ref{fig:two}(b)] and the shallow electron-like bands near M [Fig. \ref{fig:two}(d)] to the Fe 3$d$ orbitals, while the hole-like band near X [Fig. \ref{fig:two}(c)] and the deep electron-like band near M [Fig. \ref{fig:two}(e)] are attributed to the Ti 3$d$ orbitals. The Fe-3$d$ related band structure resembles those of other IBSCs. In the IBSCs, there are generally two electron-like bands from Fe 3$d$ near M. Only one electron-like band from Fe 3$d$ is observed in our experiments [Fig. \ref{fig:two}(d)], while another one from Fe 3$d$ is not identified, most likely due to its extremely low spectral weight as it is folded from the adjacent M due to inequivalent As sites around Fe. The Ti-3$d$ related band structure is similar to that of BaTi$_2$As$_2$O \cite{FengDL}. In BaTi$_{2}$As$_{2}$O, there are one electron-like band near $\Gamma$, one hole-like band near X and one electron-like band near M, which cross $E_F$. We do observe one electron-like band with a bottom of --0.5 eV at $\Gamma$ [Fig. \ref{fig:two}(b)] and our photon energy dependence measurements suggest it to originate mainly from Ti 3$d$ states [Fig. S1 in the Supplement Materials], in agreement with the observation in BaTi$_2$As$_2$O \cite{FengDL}. The spectral intensity of this band is smeared out as dispersing towards $E_F$, and its wave vector is estimated to be close to 0.4 $\pi$/$a$ along $\Gamma$M by extrapolating the band dispersion to $E_F$.

\begin{figure}[ht]
\centering
\includegraphics[scale=0.36]{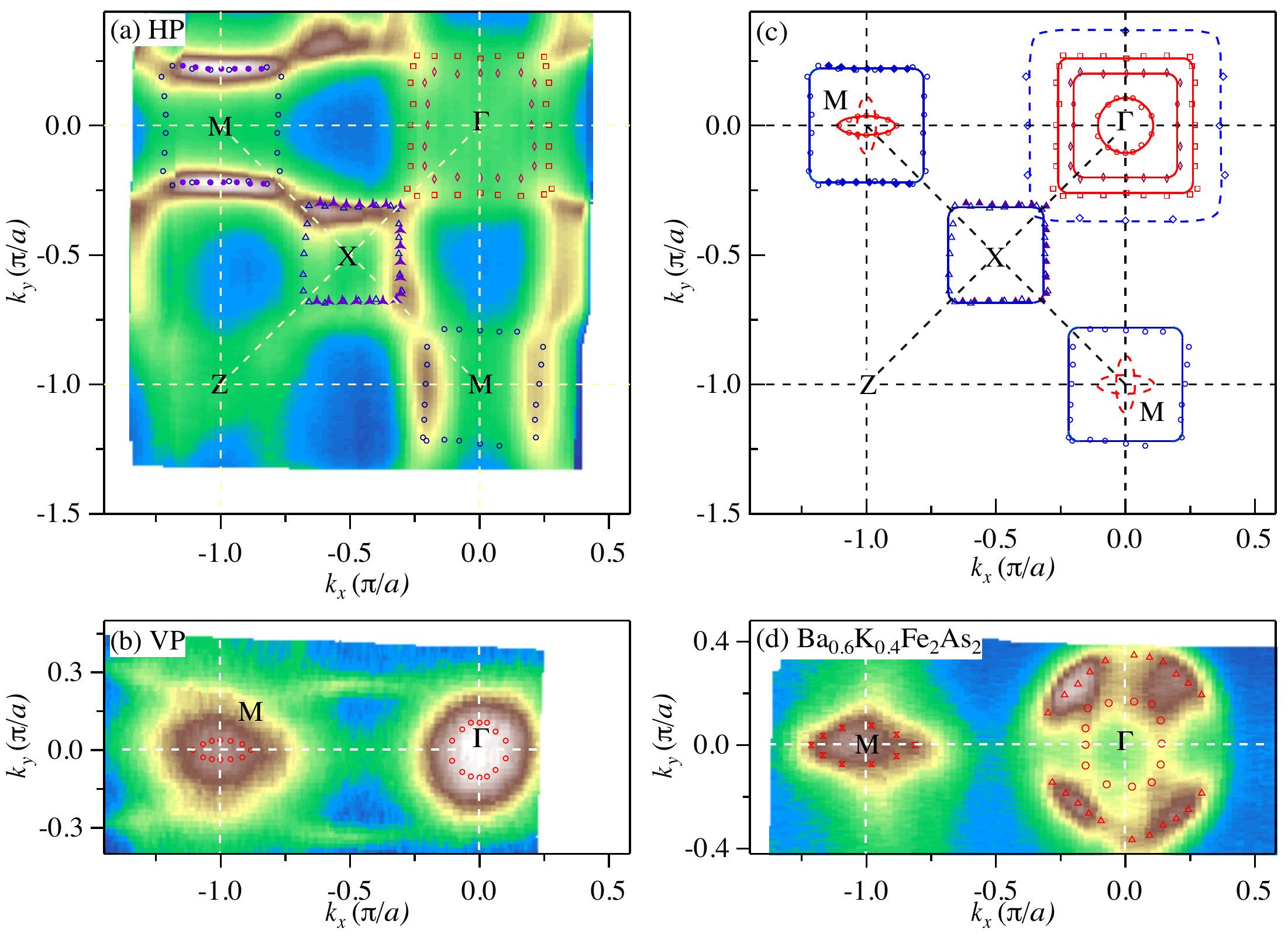}
\caption{\label{fig:three}(Color online) (a) FS intensity plot of Ba22241 recorded at 150 K and $h\nu$ = 55 eV with HP. The intensity is obtained by integrating the spectra within $\pm$ 10 meV with respect to $E_F$. $a$ represents the nearest neighbor Fe(Ti)-Fe(Ti) distance. (b) Same as (a) but taken with VP. (c) Extracted FSs of Ba22241. Red and blue symbols represent the extracted $k_F$ points related to the FeAs and Ti$_2$As$_2$O layers, respectively. Solid and hollow symbols are extracted from the data taken at 150 and 30 K, respectively. Solid curves are guides for eyes. (d) FS intensity plot of Ba$_{0.6}$K$_{0.4}$Fe$_2$As$_2$ recorded with He $I\alpha$ resonance line ($h\nu$ = 21.218 eV).}
\end{figure}

With the identification of the near-$E_F$ bands, we extract the corresponding FSs and summarize them in Fig. \ref{fig:three}(c). The extracted FSs related to the Fe $3d$ orbitals resemble those of other IBSCs, indicating that the FS topology of the FeAs layers is not changed by the intercalated Ti$_2$As$_2$O layers. One prominent feature of the Fe-related FSs is that the total enclosed area of the hole pockets near $\Gamma$ is much larger than that of the electron pockets near M. Our $k_z$ dependent measurements suggest that the electronic structure of Ba22241 is quasi-2D [Fig. S1 in the Supplement Materials]. By counting the Luttinger volume of 2D FS sheets and assuming a purely 2D FS, we obtain a hole doping of $\sim$ 0.25 per Fe site. Indeed, the Fe-related FSs look similar to those of the optimally-doped Ba$_{0.6}$K$_{0.4}$Fe$_2$As$_2$ with a hole doping of 0.2 per Fe site, as shown in Fig. \ref{fig:three}(d). Note that the inner hole pockets are almost doubly-degenerate at $k_z$ $\sim$ 0 for Ba$_{0.6}$K$_{0.4}$Fe$_2$As$_2$. By counting the volumes of the Ti-related FS sheets, we obtain an electron doping of $\sim$ 0.25 per Ti site. Both the $k_z$ and temperature dependent ARPES results reflect bulk features [Figs. S1 and S2 in the Supplement Materials], thus excluding the possibility of charge polarization on the surface. Therefore, the most significant effect of the intercalation of metallic Ti$_2$As$_2$O layers is that the electrons are transferred from the FeAs layer to the Ti$_2$As$_2$O layer. The ``washed out'' nature of the Fe-$3d$ bands is then a direct consequence of the interlayer charge transfer, since the effective hole-doping places this compound in the regime of strongly doping- and temperature-dependent coherence properties induced by Hund's rule coupling \cite{Werner NPhys}. Analyzing further the many-body self-energies obtained within our DMFT calculations we find qualitatively similar incoherent behavior for the Fe-3$d$ states in Ba22241 and optimally hole-doped BaFe$_2$As$_2$.

\begin{table*}
\caption{Orbital-resolved electron count in Wannier functions calculated by
LDA+DMFT (LDA) for Ba22241 (1st column), BaFe$_2$As$_2$ with the same
structure than in Ba22241 (2nd column), BaTi$_2$As$_2$O with the same structure than in Ba22241 (3rd column), BaFe$_2$As$_2$ (4th column) and optimally doped
BaFe$_2$As$_2$ (5th column).}

\begin{tabular}{|c|c|c|c|c|c|}
\hline
Orbital & Ba22241 & BaFe$_{2}$As$_{2}$@Ba22241 & BaTi$_{2}$As$_{2}$O@Ba22241 & BaFe$_{2}$As$_{2}$ & Ba$_{0.63}$K$_{0.37}$Fe$_{2}$As$_{2}$\tabularnewline
\hline
\hline
Fe $d$$_{z^{2}}$ & \textcolor{red}{1.30} (1.41) & 1.33 (1.42) &  & 1.35 (1.43) & \textcolor{red}{1.30} (1.43)\tabularnewline
\hline
Fe $d$$_{x^2-y^2}$ & \textcolor{red}{1.33} (1.23) & 1.34 (1.23) &  & 1.32 (1.25) & \textcolor{red}{1.32} (1.25)\tabularnewline
\hline
Fe $d_{xy}$ & \textcolor{red}{1.20} (1.40) & 1.23 (1.40) &  & 1.26 (1.40) & \textcolor{red}{1.20} (1.33)\tabularnewline
\hline
Fe $d_{xz}$+$d_{yz}$ & \textcolor{red}{1.27} (1.35) & 1.30 (1.34) &  & 1.32 (1.34) & \textcolor{red}{1.27} (1.29)\tabularnewline
\hline
Total Fe $d$ & \textcolor{red}{6.38} (6.76) & 6.50 (6.75) &  & 6.56 (6.77) & \textcolor{red}{6.36} (6.61)\tabularnewline
\hline
\hline
Ti $d$$_{z^2}$ & 0.46 (0.43) &  & 0.41 (0.42) &  & \tabularnewline
\hline
Ti $d$$_{x^2-y^2}$ & 0.39 (0.39) &  & 0.38 (0.38) &  & \tabularnewline
\hline
Ti $d_{xy}$ & 0.63 (0.69) &  & 0.60 (0.69) &  & \tabularnewline
\hline
Ti $d_{xz}$ & 0.36 (0.15) &  & 0.33 (0.15) &  & \tabularnewline
\hline
Ti $d_{yz}$ & 0.45 (0.45) &  & 0.48 (0.45) &  & \tabularnewline
\hline
Total Ti $d$ & 2.29 (2.09) &  & 2.20 (2.09) &  & \tabularnewline
\hline
\end{tabular}
\end{table*}

We analyze the electron transfer from Fe to Ti by comparing LDA and LDA + DMFT calculations for Ba22241 with experimental crystal structure, BaFe$_2$As$_2$, Ba$_{0.63}$K$_{0.37}$Fe$_2$As$_2$ and the compounds derived by splitting Ba22241 into BaTi$_2$As$_2$O and BaFe$_2$As$_2$ while keeping the same distance and angle between atoms within one layer. We construct localized Wannier-like orbitals within the same window [--8.16 eV, 8.16 eV] for all the compounds and extract the number of electrons, and the orbital-resolved electron counts for Fe-$d$ and Ti-$d$, which are displayed on Table I. For Ba$_{0.63}$K$_{0.37}$Fe$_2$As$_2$ we have taken the band structure of BaFe$_2$As$_2$ with the experimental crystal structure of optimally doped BaFe$_2$As$_2$ and calculated the chemical potential such as to obtain the correct total number of electrons. The number of electrons in the different orbitals of Ba22241 obtained from LDA is independent of the particular stacking structure of this material, as we can see by comparing the calculation on Ba22241 to calculations on BaFe$_2$As$_2$ and BaTi$_2$As$_2$O crystals. Two effects are contributing to the global hole-doping of the Fe-$d$ orbitals when going from BaFe$_2$As$_2$ to Ba22241. The effect of correlations in BaFe$_2$As$_2$ is to reduce the number of electrons in the correlated shell. If the Fe-As hybridization is weaker, electrons are more localized and this effect is enhanced. In Ba22241, the Fe-As distance is about 1$\%$ larger than in BaFe$_2$As$_2$, thus lowering the hybridization with As. The consequence can be seen by comparing the number of electrons in the Fe-$d$ Wannier orbitals in BaFe$_2$As$_2$ in the structure of Ba22241 to BaFe$_2$As$_2$ in the experimental crystal structure: There is already a loss of 0.05 electrons per Fe in Ba22241. The second effect is a transfer of electrons from Fe to Ti due to correlations. In BaTi$_2$As$_2$O, correlations tend to increase the number of electrons in the Ti-$d$ orbitals. This effect is enhanced when BaTi$_2$As$_2$O and BaFe$_2$As$_2$ are put together, leading to a transfer of about 0.1 electrons from Fe to Ti. That transfer happens notably (but not exclusively) through the Fe-$d_{z^{2}}$ -- Ti-$d_{z^{2}}$ hybridization, as Ti and Fe are on top of each other. We stress that the band with bottom around -0.5 eV at the $\Gamma$ point is of mixed Ti $d_{z^{2}}$ and Fe $d_{z^{2}}$ characters, and wrongly predicted by LDA while correctly captured by our two-correlated shells LDA + DMFT.

In conclusion, we studied the electronic structure of Ba22241 and revealed a large charge transfer between the FeAs and intercalated Ti$_2$As$_2$O layers, which is identified as a consequence of electronic correlations in the 3$d$ shells by the LDA + DMFT calculations. This provides an alternative route of effective doping without element substitution for the IBSCs and thus without introducing disorder. Furthermore, our results prove the presence of strong interlayer coupling in Ba22241 arising from the Fe-$d_{z^{2}}$ -- Ti-$d_{z^{2}}$ hybridization and the metallic nature of the intercalated Ti$_2$As$_2$O layer. It has been argued that, as in the high-$T_c$ cuprate superconductors, the FeAs interlayer coupling can play a crucial role in enhancing $T_c$ in the IBSCs \cite{Cava PANS}. Further studies for the IBSCs with metallic intercalated layers may be a fruitful path for reaching higher $T_c$ in the IBSC family.

\begin{acknowledgments}
This work was supported by grants from CAS (2010Y1JB6 and XDB07000000), MOST (2010CB923000, 2013CB921700, and 2011CBA001000), NSFC (11004232, 11234014, and 11274362), the Cai Yuanpei program, the French ANR via project PNICTIDES, IDRIS/GENCI under project 091393, the European Research Council under project 617196, and the Sino-Swiss Science and Technology Cooperation (IZLCZ2138954). This work is based in part upon research conducted at the Synchrotron Radiation Center, which is primarily funded by the University of Wisconsin-Madison with supplemental support from facility Users and the University of Wisconsin-Milwaukee.
\end{acknowledgments}

\widetext
\clearpage
\begin{center}
\textbf{\large Correlation-induced self-doping in intercalated iron-pnictide superconductor Ba$_2$Ti$_2$Fe$_2$As$_4$O - Supplementary material}
\end{center}
\twocolumngrid
\setcounter{equation}{0}
\setcounter{figure}{0}
\setcounter{table}{0}
\setcounter{page}{1}
\makeatletter
\renewcommand{\theequation}{S\arabic{equation}}
\renewcommand{\thefigure}{S\arabic{figure}}
\renewcommand{\bibnumfmt}[1]{[S#1]}
\renewcommand{\citenumfont}[1]{S#1}

\begin{figure}[ht]
\centering
\includegraphics[scale=0.52]{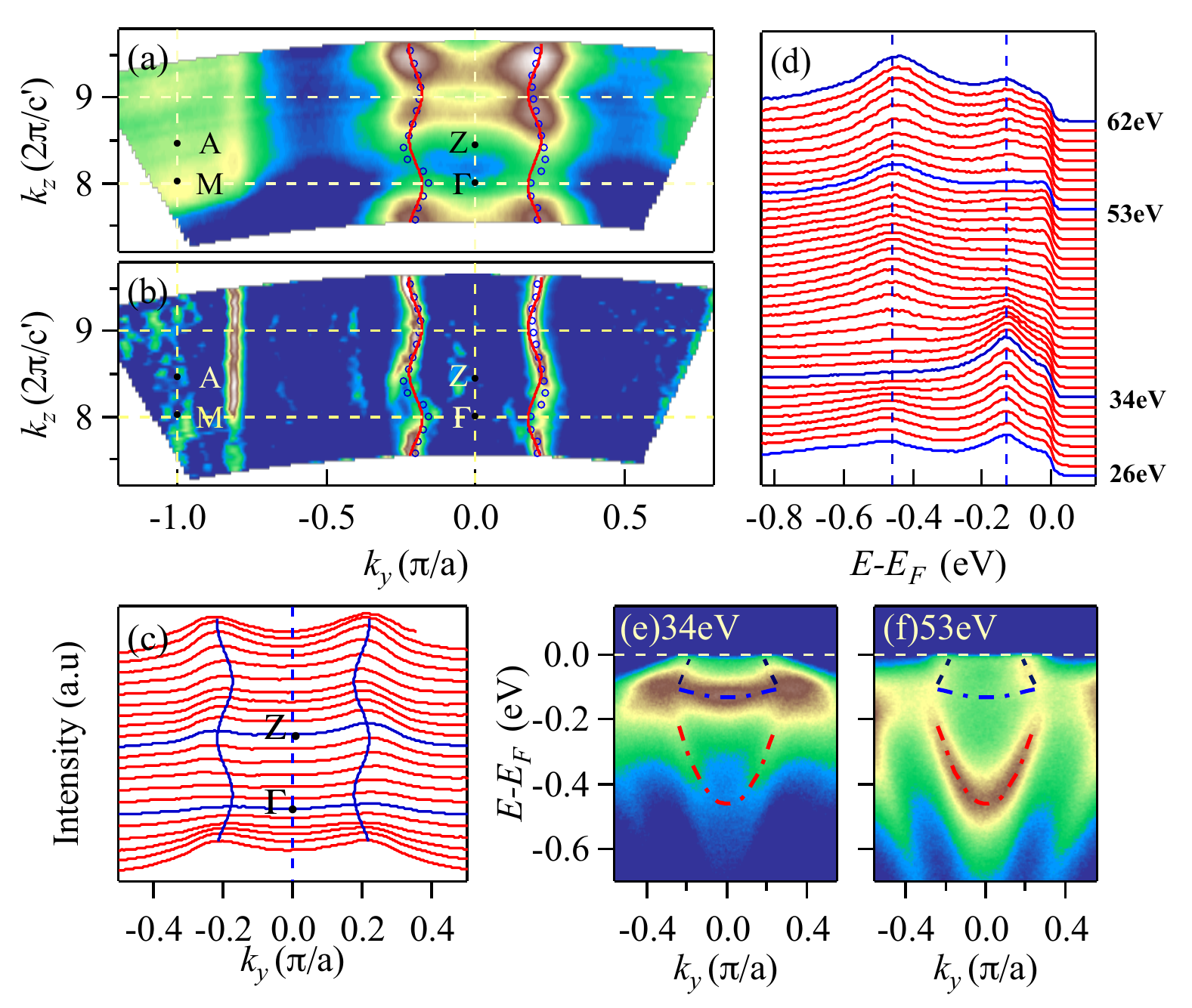}
\caption{\label{fig:kz}(a) ARPES intensity plot in the $k_y$-$k_z$ plane recorded at 25 K with different photon energies from 35 to 60 eV with horizontal polarization. The intensity is obtained by integrating the spectra within $\pm$10 meV with respect to $E_F$. An inner potential of 14 eV is used to obtain $k_z$. $c'= c/2$, where c represents the lattice parameter perpendicular to the FeAs plane. (b) Curvature intensity plot of the ARPES intensity in panel (a). (c) Corresponding momentum distribution curves (MDCs) at $E_F$ with different $k_z$. (d) Valence band spectra at the Brillouin zone center recorded with different photon energies from 26 to 62 eV. (e) and (f) ARPES intensity plots at $k_x$ = 0 recorded with 34 and 53 eV photons, respectively.}
\end{figure}

As seen in Fig. \ref{fig:kz} (a)-(c), the Fermi surface near the Brillouin zone centre exhibits an obvious warping with a period of 2$\pi/c'$, suggesting that the ARPES data reflect the bulk electronic structure. This Fermi surface is assigned to the hole-like Fe 3$d_{xz}$ band. This indicates that the interlayer coupling between the adjacent FeAs layers is not negligible, which can be attributed to the metallic nature of the intercalated BaTi$_2$As$_2$O layer.

In Fig. \ref{fig:kz} (d), the spectrum at the Brillouin zone centre has two structures, which correspond to the shallow flat band and the bottom of the deep band, respectively, in Fig.\ref{fig:kz} (e) and (f). The spectral intensities of the shallow and deep ones are strongly suppressed at $h\nu$ = 53 and 34 eV, respectively, corresponding to the Fe 3$p$-3$d$ and Ti 3$p$-3$d$ antiresonances, respectively \cite{HongD}. 

\begin{figure}[ht]
\centering
\includegraphics[scale=0.5]{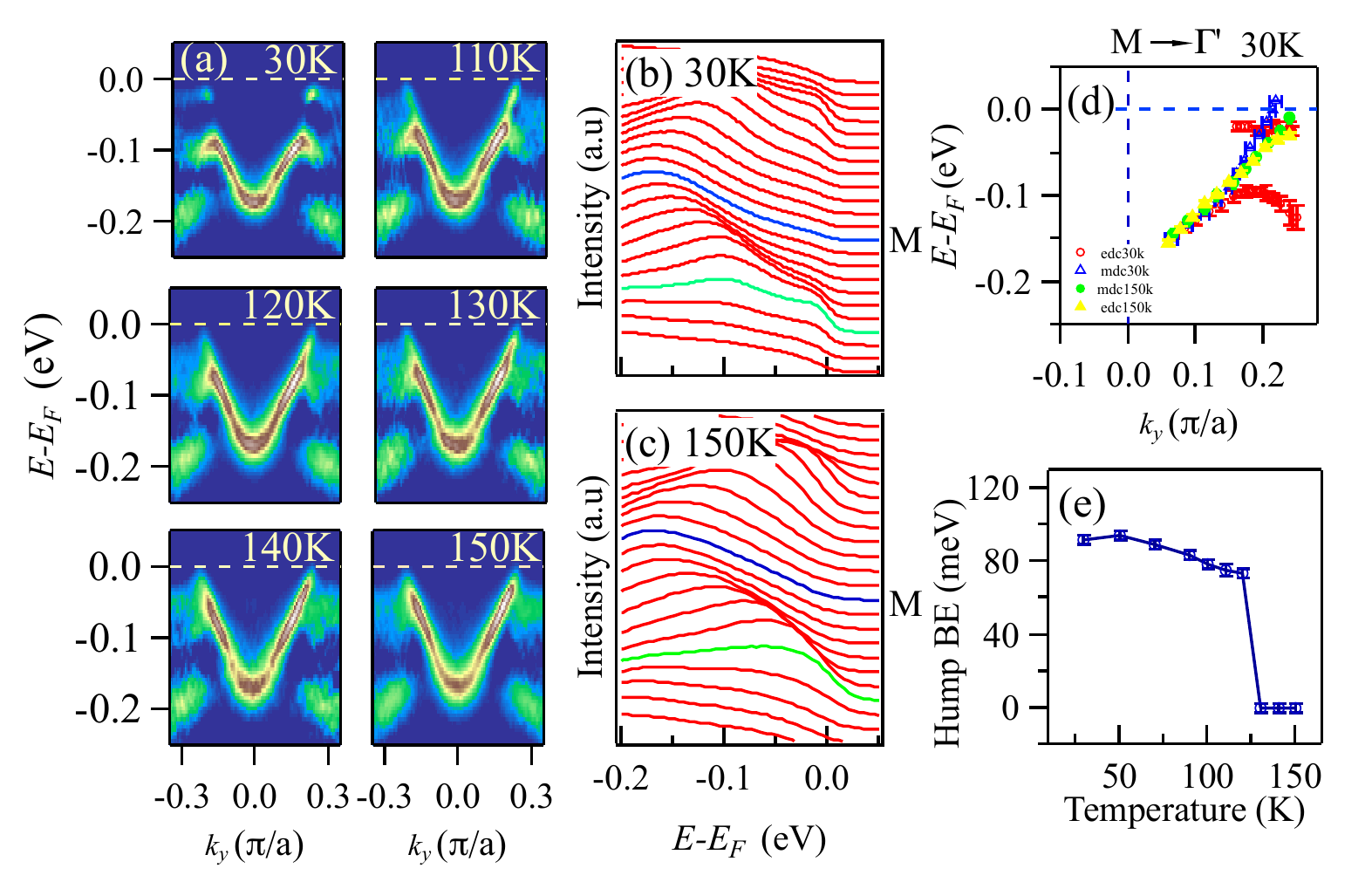}
\caption{\label{fig:T}(a) Two-dimension curvature intensity plots of the cuts at M ($k_x$ = $\pi/a$) at various temperatures between 30 and 150 K. (b) and (c) Corresponding energy distribution curves (EDCs) at 30 and 150 K, respectively. (d) Band dispersions taken from EDCs and MDCs at 30 and 150 K. (e) Binding energy of the top of the hump as a function of temperature.}
\end{figure}

The band dispersion around M derived from Ti 3$d$ exhibits a dramatic change around 120 K, which is consistent with the anomaly in resistivity and magnetic susceptibility of Ba$_2$Ti$_2$Fe$_2$As$_4$O \cite{YLSun}. Since both the $k_z$ and temperature dependent ARPES results reflect bulk features, the possibility of charge polarization on the surface can be excluded.

\begin{figure}[ht]
\centering
\includegraphics[scale=0.38]{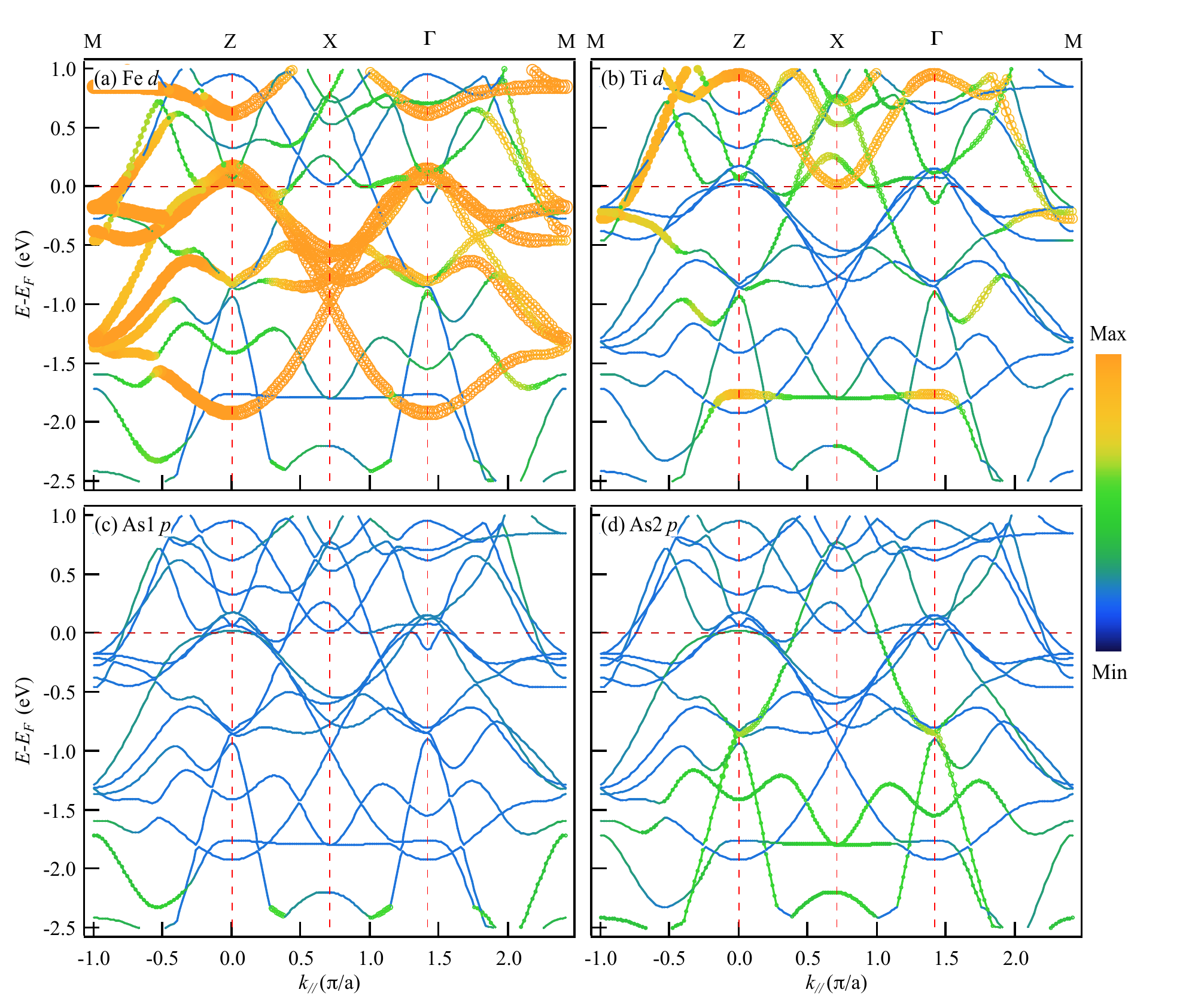}
\caption{\label{fig:LDA}(a)-(d) Contributions of the Fe 3$d$, Ti 3$d$, As1 4$p$ in the FeAs layers, and As2 4$p$ in the Ti$_2$As$_2$O layers to the LDA calculated band structure of Ba$_2$Ti$_2$Fe$_2$As$_4$O, respectively. The contribution is represented by both the symbol size and the color scale.}
\end{figure}

\end{document}